\documentclass[10pt, conference, letterpaper]{IEEEtran}    % For infocom

\def\ignore#1{}

\usepackage{amsmath,epsfig,epsf,psfrag,amssymb,amsfonts,latexsym,amsmath,color}

\usepackage{graphicx}
\usepackage{caption}
\usepackage{epstopdf}
\DeclareGraphicsExtensions{.eps}

\usepackage{bbm}
\usepackage{verbatim}
\usepackage{cite,enumerate}
\usepackage{srcltx,cases}
\usepackage[mathscr]{eucal}
\usepackage{url}
\usepackage{hyperref}
\usepackage{multirow}

\newtheorem{definition}{Definition}
\newtheorem{theorem}{Theorem}
\newtheorem{lemma}{Lemma}

\newtheorem{corollary}{Corollary}
\newtheorem{remark}{\textit{Remark}}

\newtheorem{example}{Example}

\newcommand{\beq}{\begin{equation}}
\newcommand{\eeq}{\end{equation}}

\newcommand{\upperRomannumeral}[1]{\uppercase\expandafter{\romannumeral#1}}

\allowdisplaybreaks
\graphicspath{ {./Figs/} }
\IEEEoverridecommandlockouts

\begin{document}
	\bstctlcite{BSTcontrol:compact}

	\title{ On the Effect of Shadowing Correlation on Wireless Network Performance}
	\author{Junse Lee and Fran\c{c}ois Baccelli\\
		The University of Texas at Austin, Austin, TX, USA  \\
		junselee@utexas.edu, francois.baccelli@austin.utexas.edu
		\thanks{This work was supported in part by the National Science Foundation under Grant
			No. NSF-CCF-$1218338$ and an award from the Simons Foundation $(\#197982)$, both
			to the University of Texas at Austin.}
	}

\date{}
\maketitle

\begin{abstract}
We propose and analyze a new shadowing field model meant to capture spatial correlations. The interference field associated with this new model is compared to that of the widely used independent shadowing model. Independent shadowing over links is adopted because of the resulting closed forms for performance metrics, and in spite of the well-known fact that the shadowing fields of networks are spatially correlated. The main purpose of this paper is to challenge this independent shadowing approximation. For this, we analyze the interference measured at the origin in networks where 1) nodes which are in the same cell of some random shadowing tessellation share the same shadow, or 2) nodes which share a common mother point in some cluster process share the same shadow. By leveraging stochastic comparison techniques, we give the order relation of the three main user performance metrics, namely coverage probability, Shannon throughput and local delay, under both the correlated and the independent shadowing assumptions. We show that the evaluation of the considered metrics under the independent approximation is systematically pessimistic compared to the correlated shadowing model. The improvement in each metric when adopting the correlated shadow model is quantified and shown to be quite significant.
\end{abstract}

\IEEEpeerreviewmaketitle
%%%%%%%%%%%%%%%%%%%%%%%%%%%%%%%%%%%%%%%%%%%%%%%%%%%%%%%%%%%%%%%%%%%%%%%%%%%%%%%%%%%%%%%%%%%
\section{Introduction}

\subsection{Motivation}

In wireless system level analysis, most models analyze propagation using distance-based path loss functions \cite{rappaport1996wireless,goldsmith2005wireless}. Such a modeling is justified in the free space case but does not capture real-world environments with obstacles. By introducing a shadowing term, it is possible to model the effect of obstacle blockage. This term accounts for the fact that the received signal power is strongly attenuated by obstacles on the propagation path between transmitter and receiver. For a single link, this attenuation is typically modeled by a log-normal distribution \cite{goldsmith2005wireless}, which is justified by the multiplicative blockage loss and the central limit theorem \cite{coulson1998statistical}. However, this does neither capture the fact that nearby links are often blocked by \emph{common} obstacles nor the fact that the shadowing statistics highly depend on the spatial geometry of obstacles.

Stochastic geometry has been widely studied to analyze the performance of both infrastructure (e.g., cellular) and infrastructureless (e.g., D2D) networks. The papers in this research field provide highly tractable performance evaluation results in several scenarios. Since shadowing is a significant part of wireless communication, it is important to incorporate this feature in stochastic geometric models. However, as explained above, even though the shadowing effect is spatially correlated in real networks \cite{gudmundson1991correlation}, most previous stochastic geometric models assume that shadowing is spatially independent over links. 

The main purpose of this paper is to question this independence assumption and to analyze the effect of correlated shadowing fields when using stochastic geometry. For this, we provide the Laplace transforms of the interference associated with Poisson networks under spatially correlated and independent shadowing assumptions, and prove general ordering relations between them. Using the Laplace stochastic ordering \cite{shaked1995stochastic}, we also give the ordering of some important performance metrics of the two shadowing models. Especially when the metric is coverage probability, Shannon throughput or local delay, we show that the performance metric under the independent shadowing is in fact always evaluated in a pessimistic way compared to the correlated case. 

\subsection{Related Works}

\subsubsection{Correlated Shadowing}
In real networks, shadowing fields are spatially correlated \cite{goldsmith2005wireless}. However, few generative or tractable models have been proposed to represent this correlation. Gudmundson proposed the first model of correlation \cite{gudmundson1991correlation} to model the lognormal shadowing random process between a fixed base station and a moving user by an autoregressive process with an exponentially decaying autocorrelation. As a result, the spatial dependence of shadowing can be formulated by joint Gaussian distributions. The multi-base station \cite{graziosi2002general} and multi-hop network \cite{agrawal2009correlated} cases were also considered based on similar ideas. This approach also forms the basis of the models suggested by the 3GPP\cite{access2010further} and the 802.11 standardization groups\cite{erceg2004tgn}. 

These models have shortcomings. It is hard to give a clear physical interpretation to the joint Gaussian distribution used to model spatially correlated shadowing. These models give limited intuition on large and dense wireless networks. Also, complex simulation platforms are required.

\subsubsection{Stochastic Geometry and Shadowing Models}
Over the past decades, stochastic geometric models, and most notably the planar Poisson point process (PPP) model, have become popular for the analysis of network performance in wireless communications, in both the D2D \cite{weber2010overview,baccelli2006aloha,zhang2012random,zhang2012delay} and the cellular contexts \cite{andrews2011tractable,dhillon2012modeling,dhillon2012coverage}. While an independent shadowing field can easily be incorporated into the basic models \cite{ilow1998analytic,blaszczyszyn2015wireless,bai2014analysis}, there is no known approach to combine general stochastic geometry models with correlated shadowing where links at nearby locations can be blocked by the same physical obstacles. Recently, by using a Poisson line process, correlated shadowing fields of urban networks \cite{baccelli2015correlated,zhang2015indoor} and inbuilding networks\cite{lee20163} have been analyzed in a way taking this correlation into account. However, these models use blockage-based path loss functions and fail taking the distance based-term into account.

\subsubsection{Comparison of Point Processes}

Stochastic comparison tools have been used to investigate the clustering properties among point processes by evaluation of the Ripley K function, the pair-correlation function or the empty space function \cite{chiu2013stochastic}. To quantify the impact of clustering properties among point processes, the directionally convex order on point processes \cite{blaszczyszyn2009directionally} and the properties of positive and negative association \cite{burton1985scaling,pemantle2000towards} have been proposed. This was for instance used to compare certain point processes with the Poisson point process \cite{blaszczyszyn2014comparison}. 

In this paper, we consider a new type of comparison which is that of interference fields when a shadowing random field is introduced to model the blockage effects. We compare the cases where this random field is spatially correlated or not. Other propagation effects such as reflection are not considered. % We provide the Laplace transforms of the interference field under these two assumptions and give the ordering relation between the associated Laplace transforms.

\subsection{Problem Statement and Main Contributions}

%\subsubsection{Problem Statement}

As already explained, most of the previous research papers assigned shadowing variables to links independently by using an empirical distribution \cite{goldsmith2005wireless} or based on link length when topology is incorporated \cite{ilow1998analytic,blaszczyszyn2015wireless,bai2014analysis}. We will call these models (spatially) independent shadowing models.

A typical instance of the independent shadowing model is provided in \cite{bai2014analysis}. Under this model, both the centers of blockages and the base stations are deployed as Poisson point processes. The shadowing random variable of a given link is determined by an independent Poisson random variable with mean proportional to the length of the link. %Especially when obstacles are line segments with length $l$ and density $\lambda_b$, the number of blockages for a given link with length $R$ is a Poisson random variable with mean $\frac{\lambda_b l R}{2\pi}$ \cite{bai2014analysis}. 

In contrast, in the present paper, in order to represent the spatial correlation property, we assign some shadowing value based on the obstacle topology. %The characteristic of space-based function in this paper is determined by typical user’s viewpoint. 
For example, in Fig.~\ref{fig:network_ppp}, obstacles are random segments and the plane is divided into cells in which all base stations are blocked by the same number of obstacles, when seen from the origin. Such cells are not necessarily convex but connected. Another example is depicted in Fig.~\ref{fig:network_pcp}, when the base stations form a cluster process. In this network, we assign the same shadowing random variable to the base stations which share the same mother point. In contrast to the situation of Fig.~\ref{fig:network_ppp}, even very close-by points can have different shadowing random variables. This is meant to model the situation where each cluster is located at a different altitude and has different shadowing properties. From these observations, we introduce the concept of \emph{Shadowing cell} where base stations in the same \emph{Shadowing cell} have the same shadowing random variable. Each base station should not belong to more than one \emph{Shadowing cell}.

%One example of cells is determined by the number of obstacles between any point in a cell to the typical user. Under this new shadowing field, base stations in each cell have the same shadowing random variable sampled at the centroid of the cell, and for fair comparison, each base station samples i.i.d. shadow random variable with the same law sampled at the centroid.

The main question of this paper is the comparison of the interference distribution under the correlated and the independent shadowing models in the stochastic ordering sense. To provide a fair comparison, we use the same marginal shadowing laws in both cases.  
We compute the Laplace transforms of the interference observed by the typical user which is located at the origin under these two models, and we then provide the ordering relation of the three metrics for the typical user, i.e., 1) coverage probability, 2) Shannon throughput, and 3) local delay. We show that these three metrics are completely monotone functions of the interference. From well known results on the relation between the Laplace transform ordering and completely monotone functions, we obtain the ordering relations under the two shadowing assumptions.  

For the case where base stations form either a homogeneous Poisson point process or a Mat$\acute{\mbox{e}}$rn cluster process on $\mathbb{R}^2$, we provide exact expressions for the Laplace transform of interference. These expressions are provided conditioned on the \emph{Shadowing cells}, but provide a general ordering relation of the above metrics by deconditioning. Especially, if the \emph{Shadowing cells} are Mat$\acute{\mbox{e}}$rn disks \cite{haenggi2012stochastic}, we can obtain further closed form expressions by deconditioning with respect to the \emph{Shadowing cells}. %Even though the \emph{Shadowing cells} are conditioned, there is no problem to state the ordering properties of the interference and the three main performance metrics. 

Our key findings can be summarized as follows:
\begin{itemize}
	\item We provide closed-form expressions for the Laplace transform of the interference measured at the origin under the two shadowing assumptions for some generic network examples.
	\item We investigate the Laplace transforms of interference and their ordering relationship for point processes with the same point configuration but different joint shadowing distributions. 
	\item By using the Laplace stochastic ordering and the formalism of completely monotone functions, we also give the ordering relation of the three key performance metrics under the two different shadowing assumptions. 
\end{itemize}

%%%%%%%%%%%%%%%%%%%%%%%%%%%%%%%%%%%%%%%%%%%%%%%%%%%%%%%%%%%%%%%%%%%%%%%%%%%%%%%%%%%%%%%%%%%
\section{Laplace Stochastic Ordering and Completely Monotone Functions}\label{sec:math_prem}

We first introduce some mathematical preliminaries. The following results and definitions are borrowed from \cite{alzaid1991laplace}. They will be used to investigate the ordering of the network performance metrics in the next sections.

\definition[Laplace stochastic ordering] Let $X$ and $Y$ be random variables in $\mathbb{R}^+$. $X$ is said to be less than $Y$ in the \emph{Laplace stochastic ordering} (written $X\leq_{\mbox{L}}Y$), if the Laplace transforms $\mathcal{L}_X(s)=\mathbb{E}[e^{-sX}]$ and $\mathcal{L}_Y(s)=\mathbb{E}[e^{-sY}]$ satisfy
\begin{equation}
\mathcal{L}_X(s)\geq \mathcal{L}_Y(s)~~\mbox{for all $s>0$.}
\end{equation}

\definition[Completely monotone function] A real function $f$ is called \emph{completely monotone} if all its derivatives $f^{(n)}$ exist and satisfy
\begin{equation}
(-1)^nf^{(n)}(x)\geq 0~~\mbox{for all $x$ and for all $n=0,1,2,\ldots$}
\end{equation}

\example The following functions are completely monotone:
\begin{align*}
&e^{-\alpha x},~\mbox{ for $\alpha>0$;}~~\frac{1}{(\lambda+\mu x)^{\nu}},~\mbox{ for $\lambda$, $\mu$, $\nu>0$;}\\
&\ln\left(b+\frac{c}{x+d}\right),~~~~\mbox{ for $b\geq 1$, $c,~d\geq 0$.}
\end{align*}

\remark\label{rem:monotone_examples} If $f(x)$ and $g(x)$ are completely monotone functions, so are following functions:
\begin{align*}
&af(x)+bg(x),~\mbox{with}~~a,b\geq 0,\\
&f(x)g(x),~f^{(2m)}(x),~-f^{(2m+1)}(x).
\end{align*}

The following theorem states the connection between the expectation of completely monotone functions and the Laplace transform order.

\theorem\label{theo:Lap_com_mon} $X\leq_{\mbox{L}}Y$ holds if and only if
\begin{equation}
\mathbb{E}f(X)\geq \mathbb{E}f(Y)\mbox{,}
\end{equation}
for all functions $f$ with a completely monotone derivative, for which the integral exists.

\begin{IEEEproof}
	See \cite{alzaid1991laplace}.
\end{IEEEproof}

%%%%%%%%%%%%%%%%%%%%%%%%%%%%%%%%%%%%%%%%%%%%%%%%%%%%%%%%%%%%%%%%%%%%%%%%%%%%%%%%%%%%%%%%%%%
\section{System Model}

\subsection{Network Model, Signal Model and Interference}\label{sec:network model}
In this paper, we consider a generic point process which features an infinite collection of base stations scattered on the 2-dimensional Euclidean space ($\mathbb{R}^2$). Let $\Phi=\{X_i\}$ be the point process giving the locations of base stations. This point process is assumed stationary and with intensity $\lambda$. A typical user is located at the origin ($o$) of $\mathbb{R}^2$. To circumvent technical difficulties, we assume that the serving base station of the typical user is not part of the point process, $\Phi$. %Further, let the signal power distribution from the serving base station be $f_s(\cdot)$. 

We use a path-loss model based on a distance-based power law. The received signal power at $y$ from $x$ ($x\in\mathbb{R}^2$) is 
\begin{align}\label{eq:PL}
P_{x\rightarrow y}=P_{TX}h_{xy}S_{xy} d_{xy}^{-\alpha}\mbox{,} 
\end{align}
where $P_{TX}$ is the transmit power, $h_{xy}$, $S_{xy}$ and $d_{xy}$ are the channel fading coefficient, the shadowing coefficient, and the length of the channel from $x$ to $y$, respectively. Here, $\alpha$ is the path-loss exponent. We assume $\alpha>2$. Without loss of generality, we assume $P_{TX} = 1$, since this does not affect the signal to interference power plus noise ratio (SINR) distribution after proper rescaling of the thermal noise power. The fading coefficients of different links are assumed independent and we also assume all links are subject to Rayleigh fading, which is caused by multipath reception. The channel fading coefficients are hence modeled as exponential random variables, (i.e., $h_{xy}\sim\exp(1)$). We introduce a separate term $S_{xy}$ to model the blockage effect. Since we only compute the sum interference measured at $o$, for simple notation, we use $h_{x}$, $S_x$ and $d_x$ for representing the fading coefficient, the shadowing coefficient, and the distance of channel from $x$ to $o$.

\begin{figure}
	\begin{center}
		\includegraphics[width = 2.5in,height=1.6in]{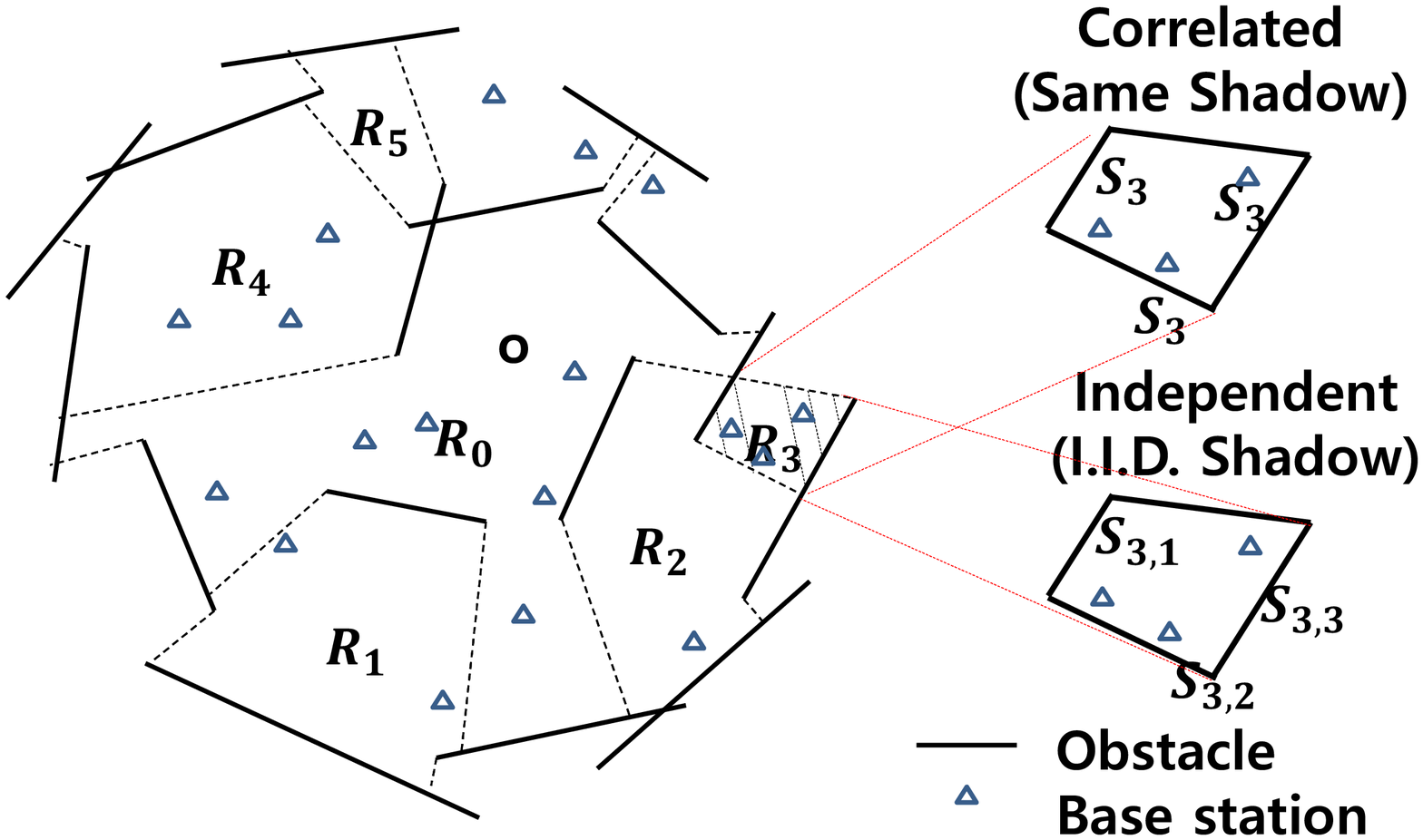}%\epsfxsize=4in {\epsfbox{Network_model.eps}}
		\caption{\small A network example with shadowing cells which are generated by line obstacles. Base stations in the same cell share a common shadowing random variable under the correlated shadowing model while base stations have i.i.d. shadowing random variable under the independent shadowing model. }
		\label{fig:network_ppp}
	\end{center}
\end{figure}

In previous wireless stochastic geometric research, the shadowing random variable, $S_x$, is either following a log-normal distribution or is a function of the length of that link which implicitly counts the number of obstacles. However, these assumptions cannot capture common blockages by the same obstacles. In contrast, the correlated shadowing model we propose is obtained by assigning the shadowing random variables in function of the obstacle topology. For example, in Fig.~\ref{fig:network_ppp}, there are 2 common blockages between all points of region $R_3$ and the origin, which defines the correlated shadowing field. One of the main simplifications in this paper is to only consider blockage and to ignore other effects such as scattering, reflection and so on. 

\begin{figure}
	\begin{center}
		\includegraphics[width = 2.5in,height=1.6in]{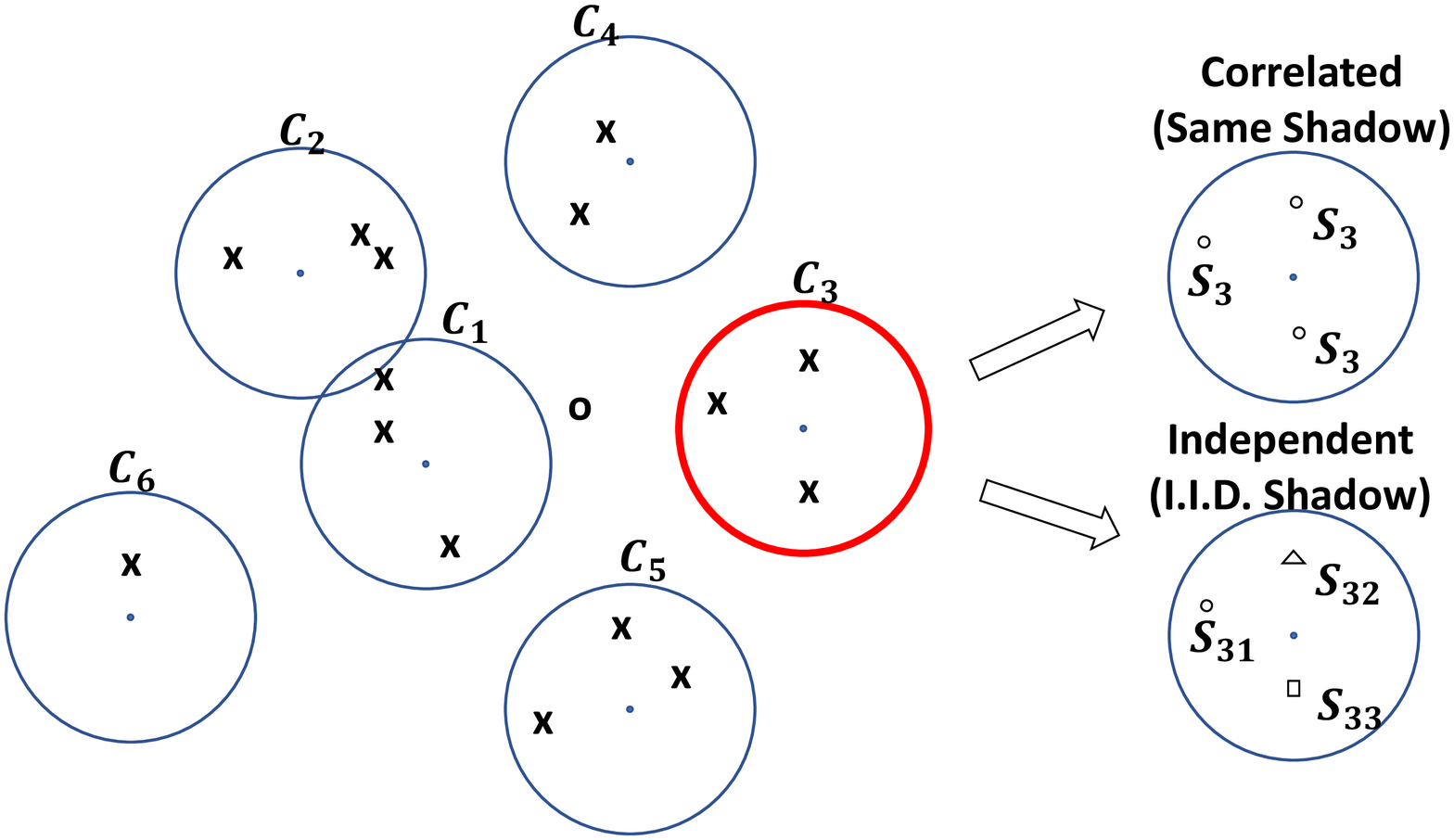}%\epsfxsize=4in {\epsfbox{Network_model.eps}}
		\caption{\small A network example based on a Mat$\acute{\mbox{e}}$rn cluster process. In this example, shadowing cells are assumed as the Mat$\acute{\mbox{e}}$rn disks.  }
		\label{fig:network_pcp}
	\end{center}
\end{figure}

Let $\{R_i\}_{i\in\mathbb{N}}$ be the set of \emph{shadowing cells} in which base stations share a common shadowing random variable to the origin. One example is illustrated in Fig.~\ref{fig:network_ppp}. In this example, the shadowing cells are determined by the position of the end points of the segments. Here, shadowing cells are connected but not necessarily convex. Another example is given in Fig.~\ref{fig:network_pcp}, where $\Phi$ is a Mat$\acute{\mbox{e}}$rn cluster process. In this example, each $R_i$ is a logical partition which is a set of daughter points sharing a common mother point. We assume that each mother point assigns a common shadowing variable to her daughter points. Further, instead of the realization of obstacles, consider a network in which the base stations with the same shadow are clustered. By assigning a shadow random variable $T_i$ which follows the distribution $f_{T_i}(\cdot)$ to all $x$ in $R_i$, namely by taking $S_x=T_i$ for all $x\in R_i$, it is possible to give different shadowing properties to different clusters. % 

For a fair comparison between the correlated and the independent shadowing fields, in both models, we use the same shadowing probability law for the points in the same shadow cell. The main difference between these two shadow assumptions is that the shadowing random variables are pathwise the same for all points in the same shadowing cell under the correlated shadowing, while they are i.i.d. under the independent shadowing model.

We provide results for general $T_i$, but for computational analysis, we consider some specific examples. For example, we will consider the case where the shadowing random variable of all base station in $R_i$ by $T_i = K^{n_i}$ where $K(<1)$ is the attenuation factor and $n_i$ is the number of obstacles between the typical user and base stations in $R_i$. Another possible scenario is that where we pick a representative point for each $R_i$ and assign the shadowing random variable as a function of the link length between this point and the origin.

Associated with a point process $\Phi$, the interference field measured at the typical user is defined as 
\begin{align*}
I(o)\triangleq \sum_{x\in\Phi}P_{x\rightarrow o}=\sum_i\sum_{x\in\Phi\cap R_i}h_{x}T_{i,x} d_{x}^{-\alpha}\mbox{,}
\end{align*}
where $T_{i,x}$ follows the distribution $f_{T_i}(\cdot)$. $T_{i,x}$ is the shadowing coefficient of $x$ seen by the typical user when $x$ is in $R_i$. Under the correlated model, $T_{i,x}=T_i$ for all $x$ in $R_i$, while under the independent model, the random variables $T_{i,x}$ are i.i.d. The main comparison results bear on the difference of the distribution of $I(o)$ under independent and correlated $S_{x}$. %In the next section, to compute the distribution of $I(o)$, we will use the Laplace transform of $I(o)$.

\subsection{SINR Distribution and Performance Metrics}

The performance metrics we consider are all related to the SINR of the typical user, which is defined as
\begin{align}\label{eq:SINR}
\mathrm{SINR}_o=\frac{P_{\delta\rightarrow o}}{\sum_i\sum_{x\in\Phi\cap R_i}h_{x}T_{i,x} d_{x}^{-\alpha}+N}\mbox{.}
\end{align}
Here, $\delta$ is the serving base station which is not included in $\Phi$ and $N$ is the thermal noise power.

In this paper, we investigate the following metrics.

\subsubsection{Coverage Probability}

The \emph{coverage probability} of the typical user is 
\begin{align}
\mathbb{P}\left[\mathrm{SINR}_o>T\right]\mbox{,}
\end{align}
where $T$ is some target SINR for reliable communication. This can be thought as 1) the probability that the SINR of a random user exceeds $T$; 2) the average fraction of the network area within reliable communication at any time. This is the complementary cumulative distribution function (CCDF) of the SINR.

\subsubsection{Shannon Throughput}

The coverage probability is an outage-based metric. However, if a transmitter can adjust its coding rate with respect to the quality of its channel, in the so called adaptive coding case, the \emph{Shannon throughput} \cite{baccelli2009stochasticvol2} is a more relevant quantity. The \emph{Shannon throughput} of the tagged user is
\begin{align}
\mathbb{E}\left[\log(1+\mathrm{SINR}_o)\right]\mbox{.}
\end{align}
This can be thought as the expectation of the bit rate of a random user in the network when adaptive coding is used.

\subsubsection{Local Delay}

We also consider a packet model. One of the important metrics in this model is the mean time to transmit a packet, which is referred as the \emph{local delay} \cite{baccelli2009stochasticvol2}. This model requires a time-space structure. 	

In this setting, the realization of the set of transmitters $\Phi$ remains unchanged over time, while channel coefficients vary over time. More precisely, for a sequence of time slots $n=1,2,\ldots$, $\Phi(n)=\Phi$, where $\Phi(n)$ is the transmitter process at time $n$ but $h_{\mathbf{xy}}(n)$ for all $\mathbf{x},\mathbf{y}\in\mathbb{R}^2$ are i.i.d. channel random variables with respect to $n$.

The local delay given $\Phi$ is defined as the first $n$ when the SINR of the typical link at the $n$-th time is larger than some threshold $T$, i.e.,
\begin{align}
L_{\Phi}=\inf\{n\geq 1:\mathrm{SINR}_o[n]\geq T|\Phi\}\mbox{.}
\end{align}
Here, $\mathrm{SINR}_o[n]$ is the typical user's SINR measured at time $n$:
\begin{align}
\mathrm{SINR}_o[n]=\frac{h_{\delta}[n]T_{\delta}d_{\delta}^{-\alpha}}{\sum_i\sum_{x\in\Phi\cap R_i}h_{x}T_{i,x} d_{x}^{-\alpha}+N}\mbox{,}
\end{align}
where $h_y[n]$, $T_y$, $d_y$ are the fading coefficient of the link between $y$ and the typical user at time $n$, the shadowing random variable of $y$ seen by the typical user, and the link length between $y$ and the typical user, respectively, where $y\in\{x\in\Phi, \delta~\mbox{(serving base station)} \}$.

%%%%%%%%%%%%%%%%%%%%%%%%%%%%%%%%%%%%%%%%%%%%%%%%%%%%%%%%%%%%%%%%%%%%%%%%%%%%%%%%%%%%%%%%%%%
\section{Interference Field of Poisson Shadowing with Common Randomness}\label{sec:inter_field}

In this section, we provide the Laplace transforms of the interference field at the origin under Poisson assumptions for both the correlated and the independent shadowing fields. 

There are several types of Poisson networks modeling both cellular and ad-hoc networks. We mainly consider two types of networks where $\Phi$ is modeled by 1) a homogeneous PPP or 2) a Poisson cluster process (PCP). These point processes are widely used to model communication networks\cite{andrews2011tractable,ganti2009interference}. We denote the interference measured at the origin under the homogeneous PPP model with the independent and the correlated shadowing by $I_{ind,p}(o)$ and $I_{cor,p}(o)$, respectively, and by $I_{ind,c}(o)$ and $I_{cor,c}(o)$ under the homogeneous PCP model. respectively.

%For all following network models, we assume the interference channels are subject to Rayleigh fading, i.e., $h_x\sim\exp(1)$.

%
\subsection{Poisson Point Process Network Model}\label{sec:model1}
%The first network model is a homogeneous Poisson point process with intensity $\lambda$ on $\mathbb{R}^2$. We denote the set of transmitters by $\Phi_p$. Instead of modeling a random blockage process with a Boolean model %as in \cite{bai2014analysis}, we randomly divide $\mathbb{R}^2$, and denote the resulting cells of the planes by $R_i$, $i\in\mathbb{N}$, with $\bigcup_i R_i=\mathbb{R}^2$. We denote the set of transmitters in $R_i$ by %$X_i=\Phi_p\cap R_i$ and the $k$-th transmitter of $X_i$ and its mark by $X_{ik}$, and $S_{ik}$, respectively. So, $X_i=\{X_{ik}\}_{k=1,\ldots,|X_i|}$ and $\bigcup_i\bigcup_{k=1}^{|X_i|} X_{ik} = \Phi_p$.

%Now, we model a mark distribution for representing the correlated and the independent shadowing fields. 

Consider the case where $\Phi$ is a homogeneous PPP.

\theorem\label{theo:PPP_Lap} When $\Phi$ is a homogeneous PPP with intensity $\lambda$ and the channels are subject to Rayleigh fading, the conditional Laplace transforms of $I(o)$ given the shadowing cells $\{R_i\}$ are
\begin{align}\label{eq:PPP_Lap_ind}
&\mathcal{L}_{I_{ind,p}(o)|\{R_i\}}(s)\nonumber\\&=\prod_i\exp\left(-\lambda  \int_{R_i}\left(1-\mathbb{E}_{T_i}\left[\frac{1}{1+s\|x\|^{-\alpha}T_i}\right]\right) dx\right)\mbox{,}
\end{align} 
under the independent shadowing assumption, and 
\begin{align}\label{eq:PPP_Lap_cor}
&\mathcal{L}_{I_{cor,p}(o)|\{R_i\}}(s)\nonumber\\&=\prod_{i}\mathbb{E}_{T_i}\left[\exp\left(-\lambda \int_{R_i}(1-\frac{1}{1+s\|x\|^{-\alpha}T_i})dx\right)\right]\mbox{,}
\end{align} 
under the correlated shadowing assumption.

\begin{IEEEproof}
	See Appendix~\ref{appen:theo_PPP_Lap}.	
	%Both $\mathcal{L}_{I_{ind,p}(o)|\{R_i\}}(s)$ and $\mathcal{L}_{I_{cor,p}(o)|\{R_i\}}(s)$ come from the result of Lemma~\ref{lem:PGFL} by conditioning and deconditioning with respect to the small scale fading coefficients $h_x$. The main difference between the two results is the expectation location. This comes from the conditioning and deconditioning over the independent and the common random variables.
\end{IEEEproof}

\corollary\label{cor:mean_model1} The mean interferences observed by the typical user under the correlated shadowing and the independent shadowing are the same.
\begin{align}\label{eq:cor_model1_1}
\mathbb{E}[I_{cor,p(o)}]=\mathbb{E}[I_{ind,p(o)}]\mbox{.}
\end{align}

The means of the interference fields under the two shadowing assumptions are the same. This substantiates the claim of fairness of the comparison between the two shadowing assumptions.

\corollary\label{cor:var_model1} The variance of the interferences observed by the typical user under the correlated shadowing and the independent shadowing satisfy
\begin{align}\label{eq:cor_model1_2}
var[I_{cor,p}(o)]\geq var[I_{ind,p}(o)]\mbox{.}
\end{align}
\begin{IEEEproof}
	The mean and variance of a random variable can be obtained by differentiating its Laplace transform at $s=0$.  
	\begin{align}
	&\mathbb{E}[X]=-\frac{d}{ds}\mathcal{L}_X(s)|_{s=0},\\
	&var[X]=\mathbb{E}[X^2]-(\mathbb{E}[X])^2\nonumber\\&=\frac{d^2}{ds^2}\mathcal{L}_X(s)|_{s=0}-\left(\frac{d}{ds}\mathcal{L}_X(s)|_{s=0}\right)^2\mbox{.}
	\end{align}
	By leveraging these relations, we can obtain the conditional mean and the variance of the interference given the shadowing cell $\{R_i\}_{i\in\mathbb{N}}$ as
	\begin{align}
	&\mathbb{E}[I_{cor,p}(o)|\{R_i\}_{i\in\mathbb{N}}]=\mathbb{E}[I_{ind,p}(o)|\{R_i\}_{i\in\mathbb{N}}]\nonumber\\
	&=\lambda\sum_{i\in\mathbb{N}}\mathbb{E}[T_i]\int_{R_i}x^{-\alpha}dx\mbox{,}\\
	&var[I_{ind,p}(o)|\{R_i\}_{i\in\mathbb{N}}]\nonumber\\
	&=2\lambda\sum_{i\in\mathbb{N}}\mathbb{E}[T_i^2]\int_{R_i}x^{-2\alpha}dx+\lambda^2\sum_{i\in\mathbb{N}}\mathbb{E}[T_i]^2\left(\int_{R_i}x^{-\alpha}dx\right)^2{,}\\
	&var[I_{cor,p}(o)|\{R_i\}_{i\in\mathbb{N}}]-var[I_{ind,p}(o)|\{R_i\}_{i\in\mathbb{N}}]\nonumber\\
	&=\lambda^2\sum_{i\in\mathbb{N}}var[T_i]\left(\int_{R_i}x^{-\alpha}dx\right)^2\geq 0\mbox{.}
	\end{align}
	We obtain \eqref{eq:cor_model1_1} and \eqref{eq:cor_model1_2} by deconditioning with respect to $\{R_i\}_{i\in\mathbb{N}}$.
\end{IEEEproof}

\remark When two random variables, $X$ and $Y$ have the same mean but $X$ has bigger variance, the Laplace transform of $X$ is bigger than that of $Y$ for very small $s$. This comes from a Taylor series expansion of $e^{-x}$, and higher order terms are neglected for small $s$. So, for small $s$, $\mathcal{L}_{I_{cor,p}}(s)\geq\mathcal{L}_{I_{ind,p}}(s)$. We can obtain the following and stronger ordering relation in the next corollary.

\theorem\label{cor:compare_lap_PPP} For all $s>0$,
\begin{align}\label{eq:cor_model1_3}
\mathcal{L}_{I_{cor,p}(o)}(s)\geq \mathcal{L}_{I_{ind,p}(o)}(s)\mbox{.}
\end{align}
\begin{IEEEproof}
	This relation simply comes from Jensen's inequality by comparing Equations~\eqref{eq:PPP_Lap_ind} and \eqref{eq:PPP_Lap_cor}. We obtain \eqref{eq:cor_model1_3} by deconditioning with respect to $\{R_i\}_{i\in\mathbb{N}^+}$.
\end{IEEEproof}

\remark\label{rem:converge_model1} From Corollary~\ref{cor:var_model1}, the difference between $var[I_{cor,p}(o)|\{R_i\}_{i\in\mathbb{N}}]$ and $var[I_{ind,p}(o)|$ $\{R_i\}_{i\in\mathbb{N}}]$ is $\lambda^2\sum_{i\in\mathbb{N}}var[T_i]\left(\int_{R_i}x^{-\alpha}dx\right)^2$. Since $var[T_i]\neq 0$, equality holds if and only if $\left(\int_{R_i}x^{-\alpha}dx\right)^2$. This is achieved when the sizes of each $R_i$ goes to zero. This relates to the fact that when all $R_i$ are very small, all points have independent shadowing random variables almost surely and the model converges to the independent model. %Then, moments of all orders become the same.

\subsection{Mat$\acute{\mbox{e}}$rn Cluster Process Network Model}\label{sec:model2}

The PCP model is also widely used in wireless communications. Especially in urban networks, users tend to move to hot spot areas and this is quite well represented by PCPs. In this model, we assume the shadowing cells are equivalent to the collection of daughter points sharing the same mother point. This is a quite reasonable assumption since the mother point can be thought as the representative point of that area. As already explained, two daughter points with different mother points may be close in the Euclidean sense and yet have different shadows; we may think of this as the situation where these transmitters are at different heights and hence have different shadowing.

There are several types of PCP models. In what follows, we mainly consider the Mat$\acute{\mbox{e}}$rn cluster process. In this model, the daughter points are uniformly located in a disk with radius $r_d$ centered on its mother point. However, the result may be generalized to other PCP models by changing the function $f(\cdot)$. The only assumption on $f(\cdot)$ is that $\int_{\mathbb{R}^2}f(x)dx<\infty$, which means the mean number of daughter points per mother point is finite. $f(\cdot)$ is used below to represent the density of the first moment measure of daughter points relative to their mother point.

We denote the density of the mother point process by $\lambda_m$, and each mother point has some daughter point process which is a nonhomogeneous point process with intensity $\frac{\lambda_d}{\pi r_d^2}$ in a disk of radius $r_d$ centered on the mother point, and $0$ outside. So, the density of the PCP is $\lambda_m\lambda_d$. 

Let $T_y$ be the shadowing random variable of the shadowing cell centered at $y$.

\theorem\label{theo:PCP_Lap}  In the Poisson cluster process with independent marks on each cluster, under independent shadowing, the Laplace transform of the interference measured at the typical user is
\begin{align}
&\mathcal{L}_{I_{ind,c}(o)}(s)\nonumber\\%\mathbb{E}\left[e^{-zI_{cor,c}}\right]\\
&=\exp\Bigg(-\lambda_m\int_{\mathbb{R}^2}\bigg[1-\exp\Big(-\lambda_d\nonumber\\&\int_{\mathbb{R}^2}\big(1- 
\mathbb{E}_{T_y} \left[\frac{1}{1+s(|x+y|)^{-\alpha}T_y}\right]          \big)f(x)dx\Big)    \bigg]dy\Bigg)\mbox{,}
\end{align}
where 
\begin{align}\label{eq:dau_dist}
f(x)=\begin{cases}
\frac{1}{r_d^2},& \text{if } \|x\|\leq r_d\\
0,              & \text{otherwise.}
\end{cases}
\end{align}
In the {Poisson cluster process with} under correlated shadowing, the Laplace transform of the interference measured at the typical point is
\begin{align}
&\mathcal{L}_{I_{cor,c}(o)}(s)\nonumber\\%\mathbb{E}\left[e^{-zI_{ind,c}}\right]\\
&=\exp\Bigg(-\lambda_m\int_{\mathbb{R}^2}\bigg[1-\mathbb{E}_{T_y}\Big[\exp\big(-\lambda_d\nonumber\\
&\int_{\mathbb{R}^2}\big(1-   \frac{1}{1+s(|x+y|)^{-\alpha}T_y}        \big)f(x)dx\big)  \Big]   \bigg]dy\Bigg)\mbox{.}
\end{align}
%where $S_y$ is a Poisson random variable with mean $\lambda_bl|y|$.

\begin{IEEEproof}
	See Appendix~\ref{appen:theo_PCP_Lap}.
	\end{IEEEproof}

\theorem As in Theorem~\ref{cor:compare_lap_PPP}, we can obtain the following ordering relation by Jensen's inequality.
\begin{align}
\mathcal{L}_{I_{cor,c}(o)}(s)\geq \mathcal{L}_{I_{ind,c}(o)}(s)\mbox{.}
\end{align}

\corollary The mean interferences observed by the typical user under the correlated shadowing and the independent shadowing are
\begin{align}
\mathbb{E}[I_{cor,c}(o)]&=\mathbb{E}[I_{ind,c}(o)]\nonumber\\&=\lambda_m\lambda_d\int_{\mathbb{R}^2}\int_{\mathbb{R}^2}\mathbb{E}[T_y]|x+y|^{-\alpha}f(x)dxdy\mbox{.}
\end{align}

\corollary The variances of the interference observed by the typical user under the correlated shadowing and the independent shadowing are
\begin{align}
var[I_{ind,c}(o)]&= 2\lambda_m\lambda_d\int_{\mathbb{R}^2}\int_{\mathbb{R}^2}\mathbb{E}_{T_y}[T_y^2]|x+y|^{-2\alpha}f(x)dxdy\nonumber\\
&+\lambda_m\lambda_d^2\int_{\mathbb{R}^2}\int_{\mathbb{R}^2}|x+y|^{-2\alpha}\mathbb{E}[T_y]^2f(x)dxdy\mbox{,}\\
var[I_{cor,c}(o)]&= var[I_{ind,c}(o)]\nonumber\\
&+\lambda_m\lambda_d^2\int_{\mathbb{R}^2}\int_{\mathbb{R}^2}|x+y|^{-2\alpha}var[T_y]f(x)dxdy\mbox{.}
\end{align}

\remark\label{rem:converge_model2} When $\lambda=\lambda_m\lambda_d$ is fixed, as $\lambda_m$ increases (or equivalently as $\lambda_d$ decreases), the variances become the same. Since as $\lambda_d$ decreases, the shadowing random variable of points become more independent and the correlated model converges to the independent one.

%%%%%%%%%%%%%%%%%%%%%%%%%%%%%%%%%%%%%%%%%%%%%%%%%%%%%%%%%%%%%%%%%%%%%%%%%%%%%%%%%%%%%%%%%%%
\section{Performance Metrics Analysis}

We are now in a position to give the ordering of the network performance metrics for the correlated and the independent shadowing field models.

\subsection{Coverage Probability}

In our network models, we assume that the serving base station of the typical user, $\delta$, is not included in $\Phi$. We assume that the distance between the typical user and $\delta$ is $d_{\text{link}}$. The links between $x\in\Phi$ and the typical user are assumed to be subject to Rayleigh fading, whereas the channel between $\delta$ and the typical user is subject to any of the following fading conditions.

\subsubsection{Rayleigh Fading} 
The instantaneous signal power is an exponential random variable with mean 1. So, the coverage probability under Rayleigh fading is
\begin{align}\label{eq:prob_ray}
\mathbb{P}[\mathrm{SINR}_o>\theta]=\mathcal{L}_{I(o)}(s)|_{s=\frac{\theta}{d_{\text{link}}^{\alpha}}}\mbox{.}
\end{align}
 
\subsubsection{Rician Fading}

Rician fading is similar to Rayleigh fading except for the existence of a dominant component. This component, for instance, can be the line-of-sight wave. When the power ratio of the dominant component over the other component, the so called Rician factor, is $\kappa$, the coverage probability is
\begin{align}\label{eq:prob_ric}
\mathbb{P}[\mathrm{SINR}_o>\theta]=e^{-\kappa}\sum_{n=0}^{\infty}\sum_{l=0}^n (-1)^{l}\frac{\kappa^n}{n!}\frac{s^l}{l!}\frac{d^l}{ds^l}\mathcal{L}_{I(o)}(s)|_{s=\frac{\theta}{d_{\text{link}}^{\alpha}}}\mbox{,}
\end{align}
which comes from the CCDF of Rician distribution.

Since the Laplace transform of a random variable $X$, $\mathcal{L}_X(s)=\mathbb{E}[e^{-sX}]$, is a completely monotone function, from Remark.~\ref{rem:monotone_examples}, the expressions in Equations \eqref{eq:prob_ray}, \eqref{eq:prob_ric} are also completely monotone functions. So, we can apply Theorem~\ref{theo:Lap_com_mon} to get the following ordering relations of coverage probabilities under these fading cases. 

\theorem\label{theo:cov_prob_rel} When the channel fading of the signaling link is Rayleigh or Rician, the coverage probability under correlated shadowing is better than under independent shadowing for all $\theta\geq 0$:
\begin{align}
\mathbb{P}[\mathrm{SINR}_{o,cor}>\theta]\geq\mathbb{P}[\mathrm{SINR}_{o,ind}>\theta]\mbox{,}
\end{align}
where $\mathrm{SINR}_{o,cor}$ and $\mathrm{SINR}_{o,ind}$ are the SINR measured at the typical user under the correlated and the independent shadowing field, respectively.

%According to the theorem in the previous section, the coverage probabilities under all cases are completely monotone functions. So, if $\mathcal{L}_{I_{cor}}(z)\geq\mathcal{L}_{I_{ind}}$ holds,
%\begin{align}
%P_{cor}[\mathsf{SINR}>\theta]\geq P_{ind}[\mathsf{SINR}>\theta]\mbox{.}
%\end{align}

\subsection{Shannon Throughput}

Shannon throughput is one more example of completely monotone functions of interference. The following ordering result then follows:
\theorem\label{theo:cov_prob_spt} The mean Shannon throughput is always larger under the correlated shadowing than under the independent shadowing, i.e.,
\begin{align}\label{eq:sha_tpt_rel}
\mathbb{E}[\log_2(1+\mathrm{SINR}_{o,cor})]\geq\mathbb{E}[\log_2(1+\mathrm{SINR}_{o,ind})]\mbox{.}
\end{align}

%\remark Since the Shannon throughput can be obtained by integration of the coverage probability, the ordering relations of Theorem~\ref{theo:cov_prob_spt} and Theorem~\ref{theo:cov_prob_rel} are preserved.

\subsection{Local Delay}

\lemma Let $I=\sum_{X_i\in\Phi}G_i/l(|X_i|)$ denote the spatial interference field measured at the typical user, where $\Phi$ is some homogeneous PPP with intensity $\lambda$ on $\mathbb{R}^2$, $\{G_i\}$ are i.i.d. random variables with Laplace transform $\mathcal{L}_G(s)$ and $l(r)$ is any response function. Let $\mathcal{L}_I(s|\Phi)=\mathbb{E}[e^{-sI}|\Phi]$ denote the conditional Laplace transform of $I$ given $\Phi$. Then,
\begin{align}
\mathbb{E}\left[\frac{1}{\mathcal{L}_I(s|\Phi)}\right]=\exp\left(-2\pi\lambda\int_{0}^{\infty}v\left(1-\frac{1}{\mathcal{L}_G(s/l(v))}\right)    dv\right)\mbox{.}
\end{align}
The proof can be found in \cite{baccelli2009stochasticvol2}.

We now give the local delay expression under fast Rayleigh fading. Fast Rayleigh fading is the case where fading coefficients are resampled at every time slot.

\theorem\label{theo:local_delay} Under the network $\Phi$ with fast Rayleigh fading, the local delay of the tagged user is
\begin{align}
\mathbb{E}[L_{\Phi}]=\exp(N\theta d_{d_{\text{link}}}^{-\alpha})\left(\mathcal{L}_{I(o)}(\theta {d_{\text{link}}}^{-\alpha})\right)^{-1}\mbox{,}
\end{align}
where $N$ is the thermal noise power, $l(\cdot)$ is the path-loss function and $d_{\text{link}}$ is the distance between the typical receiver and its associated transmitter.

\begin{IEEEproof}
	In the fast Rayleigh fading case, 
	\begin{align}
	\mathbb{P}[Fd_{\text{link}}^{-\alpha}\geq \theta (N+I(o))]=\exp(-N\theta d_{\text{link}}^{-\alpha})\mathcal{L}_{I(o)}(\theta d_{\text{link}}^{-\alpha})\mbox{,}\nonumber
	\end{align}
	where $F$ is the Rayleigh fading coefficient of the serving base station with mean 1. So, the conditional expectation of the local delay given $\Phi$ is
	\begin{align}
	\mathbb{E}[L_{\Phi}]&=\sum_{n\geq 1}\mathbb{P}[L\geq n|\Phi]\nonumber\\
	&=\sum_{n\geq 1}(1-\exp(-N\theta d_{\text{link}}^{-\alpha})\mathcal{L}_{I(o)}(\theta d_{\text{link}}^{-\alpha}))^{n-1} \nonumber\\
	&= \exp(N\theta d_{\text{link}}^{-\alpha})\left(\mathcal{L}_{I(o)}(\theta d_{\text{link}}^{-\alpha})\right)^{-1}\mbox{.}
	\end{align}\end{IEEEproof}

\corollary From Theorem.~\ref{theo:local_delay}, we can conclude that 
\begin{align}
\mathbb{E}[L_{I_{cor}(o)}]\leq\mathbb{E}[L_{I_{ind}(o)}]\mbox{,}
\end{align}
since the local delay is proportional to the inverse of the Laplace transform.

%%%%%%%%%%%%%%%%%%%%%%%%%%%%%%%%%%%%%%%%%%%%%%%%%%%%%%%%%%%%%%%%%%%%%%%%%%%%%%%%%%%%%%%%%%%

%%%%%%%%%%%%%%%%%%%%%%%%%%%%%%%%%%%%%%%%%%%%%%%%%%%%%%%%%%%%%%%%%%%%%%%%%%%%%%%%%%%%%%%%%%%
\section{Computational Results}

\begin{figure}
	\begin{center}
		\includegraphics[width = 2.35in,height=1.45in]{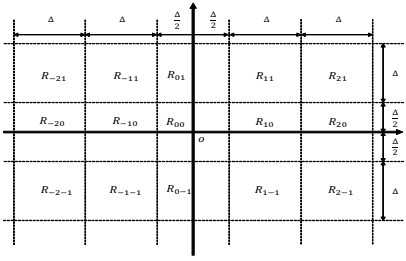}%\epsfxsize=4in {\epsfbox{grid_shadow_model1.eps}}
		\caption{\small An example of random division of $\mathbb{R}^2$. }
		\label{fig:shadow_grid}
	\end{center}
\end{figure}

\subsection{Comparison Settings}
\subsubsection{Poisson Point Process on Grid Shadowing Tessellations}\label{subsec:model1_setting}
For the shadowing tessellation case, we focus on the example of Fig.~\ref{fig:shadow_grid}, where $\{R_i\}$ consists of squares with length $\Delta$ and the origin is located at the center of one square.

We consider three cases: $\Delta=1,5,15$ with $\alpha = 4$, $\lambda=1$, $K=0.1$. For the correlated shadowing model, as $\Delta$ increases, more points will be in a square, which implies more points will share the same shadowing random variable. As $\Delta$ goes to zero, the shadowing random variables of different points become independent and the model converges to that in \cite{bai2014analysis}.

Denote the square which contains the origin by $R_{00}$, and label the other squares by their relative positions with respect to $R_{00}$ as on Fig.~\ref{fig:shadow_grid}. We denote the shadowing random variable of $R_{ij}$ by $T_{ij}$ and assume $T_{ij}=K^{r_{ij}}$, where $0\leq K\leq 1$ and $r_{ij}$ is a Poisson random variable with mean $\lambda_b\Delta\sqrt{i^2+j^2}$. Here, $K$ is the attenuation factor when signal crosses an obstacle, and $r_{ij}$ is the number of obstacles between the origin to any point in $R_{ij}$. In numerical evaluation, we assume $K=0.1$. The underlying assumptions on $r_{ij}$ are that the number of obstacles is Poisson distributed for a given link, as in \cite{bai2014analysis}; $\lambda_b$ is the implicit obstacle density and $\Delta\sqrt{i^2+j^2}$ is the distance between the center of $R_{ij}$ which is the representative point of $R_{ij}$ and the origin. Here, we assume $\lambda_b=1$. 

For the independent shadowing model, we assume the shadowing random variables of points in $R_{ij}$ are i.i.d. with distribution that of $K^{r_{ij}}$ instead of assigning one $T_{ij}$ to all points in $R_{ij}$.

Further, the distance from the typical user to its serving base station is assumed to be $0.5$ and we assume there is no obstacle between the typical user and its serving base station. We only consider the Rayleigh fading case.

\subsubsection{Mat$\acute{\mbox{e}}$rn Cluster Shadowing Cell}\label{subsec:model2_setting}

For the correlated shadowing model, we denote the common shadowing random variable of daughter points of the $i$-th mother point by $T_i$. We assume $T_i=K^{r_i}$ where $0\leq K\leq 1$ and $r_i$ is a Poisson random variable with mean $\lambda_b|X_i|$ where $|X_i|$ is the distance between $X_i$ and the origin. As in Sec.~\ref{sec:model1}, the shadowing random variable of a daughter point of the $i$-th mother point is an i.i.d. Poisson random variable with mean $\lambda_b|X_i|$. In numerical evaluation, we assume $\lambda_b = 1$ and $K=0.1$.
For the link between the typical user and its serving base station, we assume that its length is 0.5 and that it is subject to Rayleigh fading with no blockage.

In order to compare interference among networks with the same density but different amounts of the correlation, we fix the base station density by taking $\lambda_m\lambda_d=1$ and vary $\lambda_d=1,5,10$. Under correlated shadowing, as $\lambda_d$ increases, more transmitters are in the same shadowing cell and the amount of common randomness also increases. Further, we assume $r_d = 1$.

\subsection{Interpretations of Simulation Results}

\begin{figure*}[t]
	\begin{minipage}{0.32\linewidth}
		\begin{center}
			\includegraphics[width = 2.2in,height=1.4in]{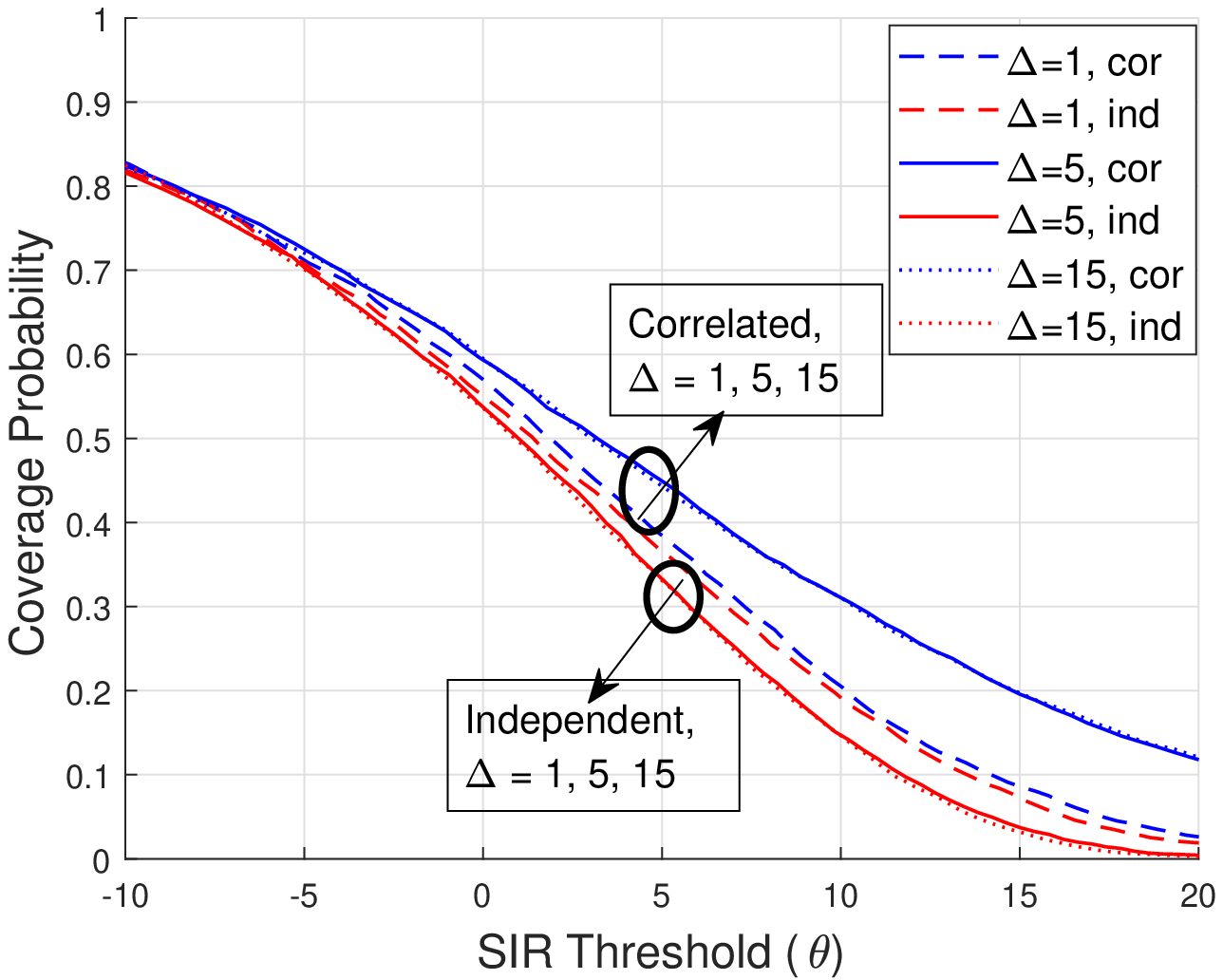}%\epsfxsize=4in {\epsfbox{grid_shadow_model1.eps}}
			\caption{\small Coverage Probability under the model in Sec.~\ref{sec:model1}. The link distance is 0.5 and $\Delta = 1,5,15$.}
			\label{fig:cov_prob_1}
		\end{center}
	\end{minipage}
	\hspace{0.01\linewidth}
	\begin{minipage}{0.32\linewidth}
		\begin{center}
			\includegraphics[width = 2.2in,height=1.4in]{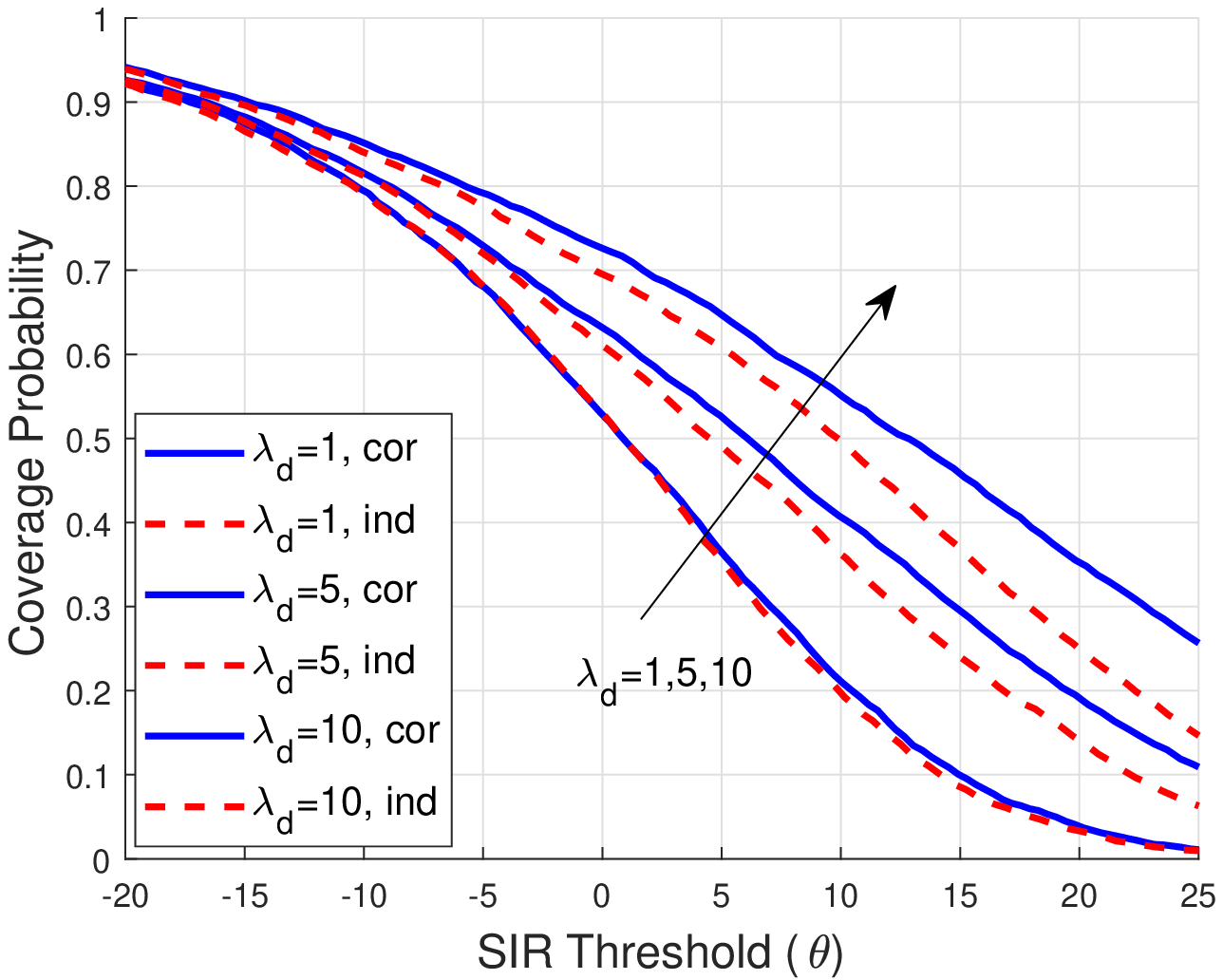}%\epsfxsize=4in {\epsfbox{grid_shadow_model2.eps}}
			\caption{\small \small Coverage Probability under the model in Sec.~\ref{sec:model2}. The link distance = 0.5 and $\lambda_d = 1,5,10$.}
			\label{fig:cov_prob_2}
		\end{center}
	\end{minipage}
	\hspace{0.01\linewidth}
	\begin{minipage}{0.32\linewidth}
		\begin{center}
			\includegraphics[width = 2.2in,height=1.4in]{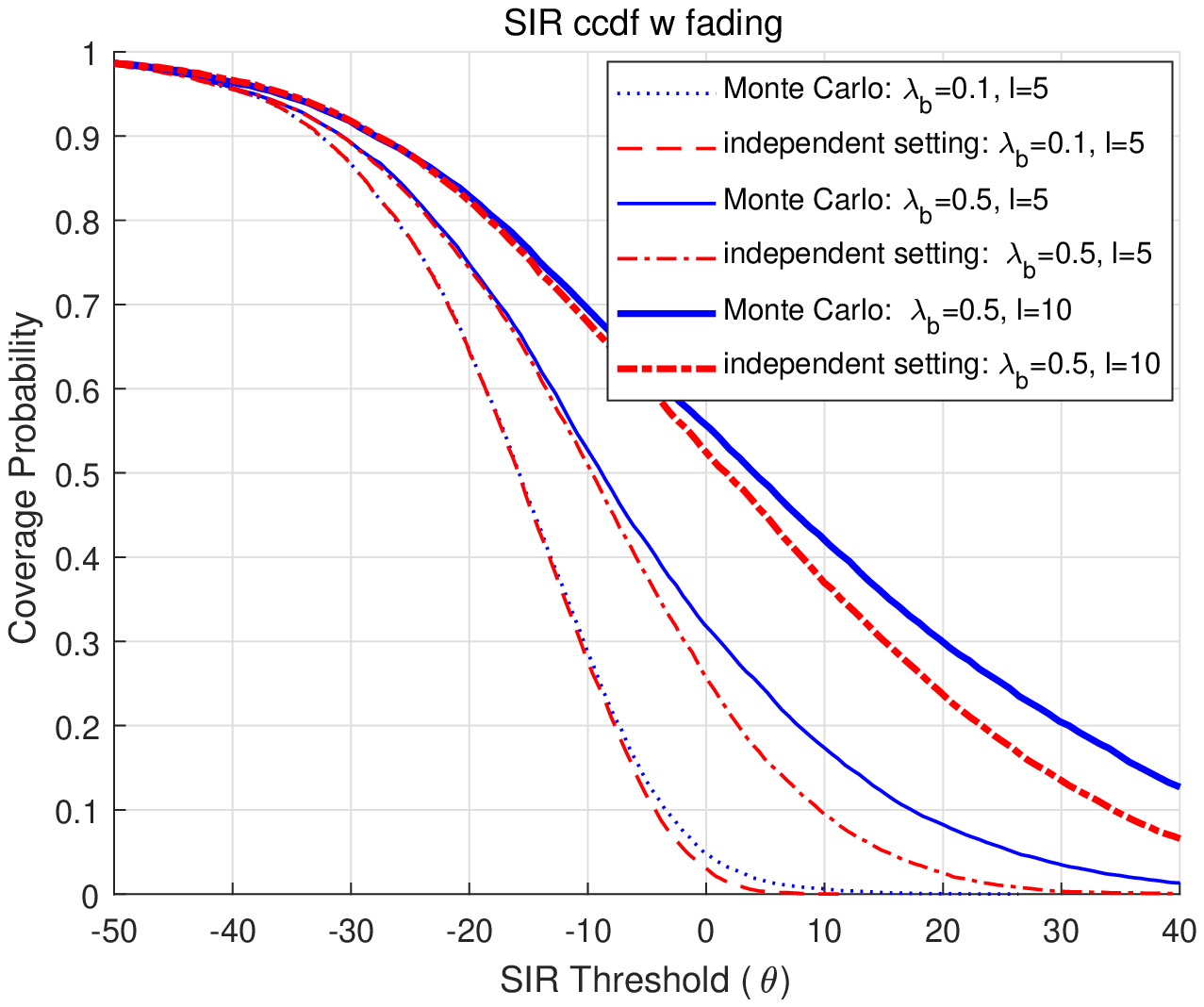}%\epsfxsize=4in {\epsfbox{grid_shadow_model2.eps}}
			\caption{\small \small Comparison of coverage probability for Fig.~\ref{fig:network_ppp} under Monte Carlo simulation and independent link assumption where $\lambda = 1$, $K=0.01$, $N=0$.}
			\label{fig:cov_prob_compare}
		\end{center}
	\end{minipage}
\end{figure*}

\subsubsection{Coverage Probability}

Fig.~\ref{fig:cov_prob_1} and \ref{fig:cov_prob_2} illustrate the coverage probabilities under the model in Sec.~\ref{subsec:model1_setting} and Sec.~\ref{subsec:model2_setting}, respectively. We can observe two facts from these simulation results which are in line with or extending our mathematical derivations:
\begin{itemize}
	\item the coverage probability under the correlated shadowing is larger than that under the independent shadowing, which confirm what is proven in Theorem.~\ref{theo:cov_prob_rel}, and
	\item the gap of coverage probabilities between correlated and independent settings decreases as $\Delta$ and $\lambda_d$ decrease.
\end{itemize}

%In Fig.~\ref{fig:cov_prob_1}, the coverage probabilities when $d=1$ for the correlated and the independent cases are close since when $\lambda=1$, the mean number of point on each $R_{ij}$ is 1 and the correlated model is almost the same as in the independent model. The discrepancy is reduced as $d$ decreases. As $d$ goes to zero, this model converges to the model in \cite{bai2014analysis} which is totally independent model over link without common randomness. 

% Also, in Fig.~\ref{fig:cov_prob_2}, the discrepancy between the correlated and the independent models becomes larger as $\lambda_d$ increases, which means there are more points which share the same shadowing random variable.

\remark In Remark \ref{rem:converge_model1} and \ref{rem:converge_model2}, we showed that the variance of the correlated model converges to that of the independent model as the sizes of all $R_i$ or $\lambda_m$ decrease. From Fig.~\ref{fig:cov_prob_1} and \ref{fig:cov_prob_2}, we can observe more general results about the convergence of higher order moments of interference between the two shadowing assumptions. The coverage probability is proportional to the Laplace transform with some positive argument when the link between the typical user and the serving base station is subject to Rayleigh fading, since 
\begin{align*}
\mathbb{P}[\frac{hd_{\text{link}}^{-\alpha}}{I+N}>T]=e^{-{d^{\alpha}_{\text{link}}}N}\times\mathcal{L}_{I}(s)|_{s=Td_{\text{link}}^{\alpha}}\mbox{,}
\end{align*}
where $h\sim\exp(1)$ is the channel fading coefficient. In Fig.~\ref{fig:cov_prob_1} and \ref{fig:cov_prob_2}, the gap of coverage probability between the two shadowing assumptions is reduced as $\Delta$ (equivalently $R_i$) decreases or $\lambda_d$ decreases. From these experimental results, we can see that the Laplace transforms (or equivalently higher order moments) of the interference of the two shadowing models become asymptotically similar.

%As $d$ which is the size of a shadowing cell decreases in Fig.~\ref{fig:cov_prob_1} or as $\lambda_d$ decreases in Fig.~\ref{fig:cov_prob_2}, the gap between the coverage probability under the correlated case and that under the independent shadowing also decreases.  This result comes from the amount of common randomness. 

Now, we provide the coverage probability results of the well known network example which is discussed in \cite{bai2014analysis}. The base stations form a homogeneous PPP with a density $\lambda$, and the obstacles are represented by a Boolean model \cite{baccelli2009stochastic}, where the centers of obstacles are distributed as a homogeneous PPP with intensity $\lambda_b$. We assume the obstacles are line segments with length $l$. See Fig.~\ref{fig:network_ppp}. For this network, we use the path loss model presented in \eqref{eq:PL} under both the correlated and the independent shadowing. Further, we assume the shadowing random variable of base station $x\in\Phi$ seen by the typical user is $K^{N_x}$, where $0<K<1$ is the attenuation factor and $N_x$ is the number of obstacles on the path between $x$ and the typical user.

In Fig.~\ref{fig:cov_prob_compare}, we give the simulation results on the independent setting and the Monte Carlo simulation of the ground truth. Under the independent setting, $N_x$ is distributed as a Poisson random variable with a mean $\frac{\lambda_b\times \text{(link length of $\overline{ox}$})}{2\pi}$. Under the Monte Carlo simulation, we deploy all base stations and obstacles, and count $N_x$ for each $x\in\Phi$ of the realized network. For both cases, we assume $\lambda=1$, $K=0.01$, $N=0$ and consider the cases with $(\lambda_b,l)=(0.1,5)$, $(0.5,5)$, and $(0.5,10)$, and assume all channels are subject to Rayleigh fading. Further, we assume $d_{\text{link}}=0.5$ with no obstacles. 

Again, from Fig.~\ref{fig:cov_prob_compare}, we can see that the coverage probability under the correlated model is better than that under the independent approximation. Also, as $\lambda_b$ or $l$ increase, the coverage probability becomes larger since more interference power is blocked by obstacles. For a quantitative result,
we pick the case with $(\lambda_b,l) = (0.5,5)$. In this case, the Monte Carlo simulation of the ground truth compared to the independent approximation is $23\%$ better at 0dB and $79\%$ better at 10dB.

\subsubsection{Shannon Throughput}

We summarize the result for Shannon throughput under the models of Sec.~\ref{subsec:model1_setting} and ~\ref{subsec:model2_setting} in Table.~\upperRomannumeral{1}. The network setting is the same as that of Fig.~\ref{fig:cov_prob_1} and Fig.~\ref{fig:cov_prob_2}. For both models, the mean Shannon throughput under the correlated shadowing is larger than that under the independent approximation as shown in Theorem.~\ref{theo:cov_prob_spt}. As for the coverage probability, as $\Delta$ and $\lambda_d$ decrease, the Shannon throughput of the correlated case converges to that of the independent approximation. Under the model in Sec.~\ref{subsec:model1_setting}, the mean Shannon rate of the correlated case compared to that of the independent approximation is from $2\%$ to $72\%$ better as $\Delta$ increases from 1 to 15, and under the model in Sec.~\ref{subsec:model2_setting}, the benefits of the correlated case over the independent approximation ranges from $6\%$ to $27\%$ as $\lambda_d$ increases from 1 to 10.

\begin{table}\label{tab:Sha_tpt}
	\begin{center}
		\begin{tabular}{|c|c|c|c|c|c|}
			\hline
			\multicolumn{3}{|c|}{PPP}&\multicolumn{3}{|c|}{PCP}\\
			\hline
			$\Delta$ & cor & ind & $\lambda_m\lambda_d=1$& cor & ind\\
			\hline
			1 &  1.9370   &  1.8942   &  $\lambda_d=1$   &  2.0180     &  1.9075  \\
			\hline
			5 &  2.7300   &   1.6358  &  $\lambda_d=5$   &   3.5615    &   2.9402 \\
			\hline
			15 &  2.7737   &   1.6101  &  $\lambda_d=10$   &    5.2018   &  4.1043  \\
			\hline
		\end{tabular}
	\end{center}
	\caption{Shannon Throughput}
\end{table}

\subsubsection{Local Delay}

\begin{figure}
		\begin{center}
			\includegraphics[width = 2.2in,height=1.4in]{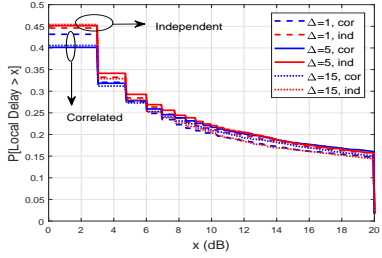}%\epsfxsize=4in {\epsfbox{grid_shadow_model2.eps}}
			\caption{\small Local delay under the model in Sec.~\ref{sec:model1}.}
			\label{fig:loc_delay_1}
		\end{center}
	\end{figure}

\begin{figure}
		\begin{center}
			\includegraphics[width = 2.2in,height=1.4in]{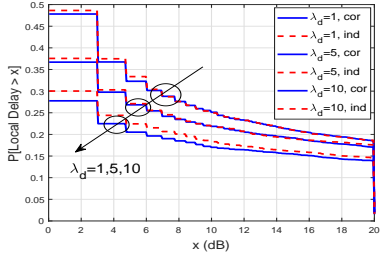}%\epsfxsize=4in {\epsfbox{grid_shadow_model2.eps}}
			\caption{\small Local delay under the model in Sec.~\ref{sec:model2}.}
			\label{fig:loc_delay_2}
		\end{center}
	\end{figure}

Fig.~\ref{fig:loc_delay_1} and Fig.~\ref{fig:loc_delay_2} illustrate the local delays under the models introduced in Sec.~\ref{sec:model1} and Sec.~\ref{sec:model2}. In both figures, the $x$-axis is in dB scale for a better visualization, and the $y$-axis is the probability that the local delay is larger than $x$. For example, in Fig.~\ref{fig:loc_delay_1}, when $\Delta=5$ under the independent case, the probability that the local delay is larger than $1$ is 0.45. From this interpretation, we can see that the local delay under the correlated case is less than that under the independent approximation. 
The probabilities that the local delay is larger than 1 under the correlated case are from $3.3\%$ to $11.7\%$ below those under the independent approximation when $\Delta = 1,15$ in Fig.~\ref{fig:loc_delay_1} and from $1.7\%$ to $7.6\%$ when $\lambda_d=1,10$ in Fig.~\ref{fig:loc_delay_2}. 

For both figures, we can see that the probabilities that the local delay exceeds $100$ are nonzero. We consider that if a network with a given topology cannot succeed in transmitting a packet in 100 time slots, it is possible to transmit packets in that topology.


%%%%%%%%%%%%%%%%%%%%%%%%%%%%%%%%%%%%%%%%%%%%%%%%%%%%%%%%%%%%%%%%%%%%%%%%%%%%%%%%%%%%%%%%%%%
\section{Conclusions}

In this paper, we analyzed the impact of correlated shadowing fields using stochastic geometry and stochastic ordering. We have shown that the Laplace transform of the interference experienced by the typical user is always larger than under the independent approximation. From this Laplace stochastic ordering, we could derive further ordering results on network performance. The main result is that when ignoring the spatial correlations of shadowing, widely used metrics such as coverage probability, Shannon throughput, local delay are systematically evaluated in a pessimistic way. For better understanding this physical phenomenon, we have provided two network examples where we could derive the exact Laplace transform of the interference distribution under the correlated and independent assumptions. By using the fact that the three key metrics are completely monotone functions, we could also prove ordering results on these metrics under the two shadowing settings. %To understand in numerical sense, we gave two experimental examples which can represent the real network scenarios with spatial correlated shadowing.

%%%%%%%%%%%%%%%%%%%%%%%%%%%%%%%%%%%%%%%%%%%%%%%%%%%%%%%%%%%%%%%%%%%%%%%%%%%%%%%%%%%%%%%%%%%
%%%%%%%%%%%%%%%%%%%%%%%%%%%%%%%%%%%%%%%%%%%%%%%%%%%%%%%%%%%%%%%%%%%%%%%%%%%%%%%%%%%%%%%%%%%

\appendices 

\section{Probability Generating Functional of a Poisson Point Process}\label{appen:preliminaries}
We first give a classical lemma for the a probability generating functional (PGFL)\cite{haenggi2012stochastic} of a Poisson point process (PPP).

\lemma\label{lem:PGFL} Let $\Phi$ be a homogeneous PPP with intensity $\lambda$ on $\mathbb{R}^2$ and $f(\cdot):\mathbb{R}^2\rightarrow [0,1]$. Then, 
\begin{align}\label{eq:PGFL}
\mathbb{E}\left[\prod_{X_i\in\Phi}f(X_i)\right]=\exp\left(-\lambda\int_{\mathbb{R}^2}(1-f(x))dx\right)\mbox{.}
\end{align} 
In Equation \eqref{eq:PGFL}, $X_i$ denotes both point $X_i$ and its coordinate. 

For the proofs in the next chapter, we state the definition of an \emph{independent marking} of a point process \cite{baccelli2009stochastic} and a lemma.

\definition[Independently marked point process]  A marked point process is said to be independently marked if, given the location of the points, the marks are mutually independent random vectors, and if the conditional distribution of mark $M_i$ of a point $x_i\in\Phi$ depends only on the location of $x_i$ it is attached to; i.e., $\mathbb{P}[M_i\in A|\Phi]=\mathbb{P}[M_i\in A|x_i] = F_{x_i}(dM)$ for some probability kernel function.

\lemma\label{lem:PGFL_ext} For an independently marked homogeneous PPP with density $\lambda$ on $\mathbb{R}^2$ and marks with distribution $F_x(dm)$ on $\mathbb{R}^l$, its Laplace transform is 
\begin{align}
\mathcal{L}_{\Phi}(f)&=\mathbb{E}[e^{-\sum_{i}\bar{f}(x_i,M_i)}]\nonumber\\
&=\exp\left[-\lambda\int_{\mathbb{R}^2}\left(1-\int_{\mathbb{R}^l}e^{-\bar{f}(x,M)}F_x(dM)\right)\right]\mbox{,}
\end{align}
for all functions $\bar{f}:\mathbb{R}^{2+l}\rightarrow \mathbb{R}^+$.

From Lemma~\ref{lem:PGFL} and \ref{lem:PGFL_ext}, we can obtain the Laplace transform of the interference fields of Poisson networks. There are several types of Poisson networks modeling both cellular and ad-hoc networks. We mainly consider two types of networks where $\Phi$ is modeled by 1) a homogeneous PPP or 2) a Poisson cluster process (PCP). These are widely used point processes to model communication networks. We denote the interference measured at the origin under the homogeneous PPP model with the independent and the correlated shadowing by $I_{ind,p}(o)$ and $I_{cor,p}(o)$, respectively, and by $I_{ind,c}(o)$ and $I_{cor,c}(o)$ under the homogeneous PCP model. respectively.

\section{Proof of Theorem \ref{theo:PPP_Lap}}\label{appen:theo_PPP_Lap}

When $\Phi$ is a homogeneous PPP with intensity $\lambda$, the conditional Laplace transform of $I_{ind,p}(o)$ given $\{R_i\}$ is
\begin{align}
\mathcal{L}_{I_{ind,p}(o)|\{R_i\}}(s)&=\mathbb{E}[e^{-sI_{ind,p}(o)}|\{R_i\}] \nonumber\\&= \mathbb{E}[e^{-s\sum_{x\in\Phi}h_xS_xd_x^{-\alpha}}|\{R_i\}]
\nonumber\\&= \mathbb{E}[e^{-s\sum_{i}\sum_{x\in\Phi\cap R_i}h_xT_id_x^{-\alpha}}|\{R_i\}]\mbox{.}\nonumber
\end{align}
By deconditioning $h_x\sim\exp(1)$ and using Lemma \ref{lem:PGFL_ext}, the conditional Laplace transform of $I_{ind,p}(o)$ given $\{R_i\}$ becomes
\begin{align}\label{eq:ppp_lap_appen_eq1}
&\mathcal{L}_{I_{ind,p}(o)|\{R_i\}}(s)\nonumber\\ &= \exp\left(-\lambda \sum_i \int_{R_i}\left(1-\mathbb{E}_{T_i,h_x}\left[e^{-sh_x\|x\|^{-\alpha}T_i}\right]\right) dx\right)\nonumber\\
& = \exp\left(-\lambda \sum_i \int_{R_i}\left(1-\mathbb{E}_{T_i}\left[\frac{1}{1+s\|x\|^{-\alpha}T_i}\right]\right) dx\right)\nonumber\\
& = \prod_i\exp\left(-\lambda  \int_{R_i}\left(1-\mathbb{E}_{T_i}\left[\frac{1}{1+s\|x\|^{-\alpha}T_i}\right]\right) dx\right)\mbox{.}
\end{align}
In the same manner, we can obtain the Laplace transform of $I_{cor,p}(o)$ as
\begin{align}\label{eq:ppp_lap_appen_eq2}
&\mathcal{L}_{I_{cor,p}(o)|\{R_i\}}(s)\nonumber\\
& = \prod_{i}\mathbb{E}_{T_i}\left[e^{-s\sum_{x\in\Phi\cap R_i}h_xT_id_x^{-\alpha}}|\{R_i\}\right]\nonumber\\
& = \prod_{i}\mathbb{E}_{T_i}\left[\exp\left(-\lambda \int_{R_i}(1-\mathbb{E}_{h_x}[e^{-sh_x\|x\|^{-\alpha}T_i}])dx\right)|\{R_i\}\right]\nonumber\\
& = \prod_{i}\mathbb{E}_{T_i}\left[\exp\left(-\lambda \int_{R_i}(1-\frac{1}{1+s\|x\|^{-\alpha}T_i})dx\right)\right] \mbox{.}
\end{align}

The main difference between \eqref{eq:ppp_lap_appen_eq1} and \eqref{eq:ppp_lap_appen_eq2} is the location of the expectation, $\mathbb{E}_{T_i}$.

\section{Proof of Theorem \ref{theo:PCP_Lap}}\label{appen:theo_PCP_Lap}

For a given mother point, its daughter points and their common shadowing random variable (for the correlated case) or their i.i.d. shadowing random variables are independent mark of the mother point. So, by applying Lemma~\ref{lem:PGFL_ext}, we can obtain the Laplace transform of interference.

Let $\Phi$ be the Mat$\acute{\mbox{e}}$rn cluster process with a Mat$\acute{\mbox{e}}$rn radius $R_d$, the intensity of mother Poisson point process $\lambda_m$, and the mean number of daughter points per mother $\lambda_d$. The generating functional of Mat$\acute{\mbox{e}}$rn process is also given by \cite{chiu2013stochastic,daley2007introduction}
\begin{align}\label{eq:PCP_PGFL}
&\mathbb{E}[\prod_{x\in\Phi}g(x)]\nonumber\\
&=\exp\left( -\lambda_m\int_{\mathbb{R}^2}\left[1-M\left(\int_{\mathbb{R}^2}g(x+y)f(y)dy \right)        \right] dx    \right)\mbox{,}
\end{align}
where $M(z)=\exp(-\lambda_d(1-z))$ and $f(\cdot)$ is in \eqref{eq:dau_dist}.

Under the independent shadowing field, the Laplace transform of $I_{ind,c}(o)$ is
\begin{align}
&\mathcal{L}_{I_{ind,c}(o)}(s)=\mathbb{E}[e^{-sI_{ind,c}(o)}]=\mathbb{E}[\prod_{x\in\Phi}e^{-sh_xS_x\|x\|^{-\alpha}}]\nonumber\\
&\stackrel{(a)}{=}\exp\Bigg( -\lambda_m\int_{\mathbb{R}^2}\Bigg[1-M\Big(\nonumber\\
&~~~~~~~~~~~~~~~~~~\mathbb{E}_{h_x,T_y}[\int_{\mathbb{R}^2}h_xT_y\|x+y\|^{-\alpha}f(y)dy] \Big)        \Bigg] dx    \Bigg)\nonumber\\
&\stackrel{(b)}{=}\exp\Bigg(-\lambda_m\int_{\mathbb{R}^2}\Big[1-\exp\bigg(-\lambda_d\int_{\mathbb{R}^2}\nonumber\\
&~~~~\left(1- \mathbb{E}_{T_y} \left[\frac{1}{1+s(|x+y|)^{-\alpha}T_y}\right]          \right)f(x)dx\bigg)    \Big]dy\Bigg)\mbox{.}
\end{align}
where (a) comes from Lemma \ref{lem:PGFL_ext} and \eqref{eq:PCP_PGFL}, and (b) is by deconditioning the channel fading coefficient $h_x$ and the shadowing random variable $T_y$. In the same manner, the Laplace transform of $I_{cor,c}(o)$ is obtained as
\begin{align}
&\mathcal{L}_{I_{cor,c}(o)}(s)\nonumber\\
&=\exp\Bigg(-\lambda_m\int_{\mathbb{R}^2}\Big[1-\mathbb{E}_{T_y}[\exp\bigg(-\lambda_d\nonumber\\
&~~~~~~~~\int_{\mathbb{R}^2}\left(1-    \frac{1}{1+s(|x+y|)^{-\alpha}T_y}        \right)f(x)dx\bigg)  ]   \Big]dy\Bigg)\mbox{.}
\end{align}
As in Theorem \ref{theo:PPP_Lap}, the main difference is the location of the expectation of $T_y$ induced by common and independent random variables.

%%%%%%%%%%%%%%%%%%%%%%%%%%%%%%%%%%%%%%%%%%%%%%%%%%%%%%%%%%%%%%%%%%%%%%%%%%%%%%%%%%%%%%%%%%%

%%%%%%%%%%%%%%%%%%%%%%%%%%%%%%%%%%%%%%%%%%%%%%%%%%%%%%%%%%%%%%%%%%%%%%%%%%%%%%%%%%%%%%%%%%%

\bibliographystyle{Bibs/ieeetran}
\bibliography{Bibs/IEEEabrv,Bibs/referenceBibs}

\begin{thebibliography}{10}
\providecommand{\url}[1]{#1}
\csname url@samestyle\endcsname
\providecommand{\newblock}{\relax}
\providecommand{\bibinfo}[2]{#2}
\providecommand{\BIBentrySTDinterwordspacing}{\spaceskip=0pt\relax}
\providecommand{\BIBentryALTinterwordstretchfactor}{4}
\providecommand{\BIBentryALTinterwordspacing}{\spaceskip=\fontdimen2\font plus
\BIBentryALTinterwordstretchfactor\fontdimen3\font minus
  \fontdimen4\font\relax}
\providecommand{\BIBforeignlanguage}[2]{{%
\expandafter\ifx\csname l@#1\endcsname\relax
\typeout{** WARNING: IEEEtran.bst: No hyphenation pattern has been}%
\typeout{** loaded for the language `#1'. Using the pattern for}%
\typeout{** the default language instead.}%
\else
\language=\csname l@#1\endcsname
\fi
#2}}
\providecommand{\BIBdecl}{\relax}
\BIBdecl

\bibitem{rappaport1996wireless}
T.~S. Rappaport \emph{et~al.}, \emph{Wireless communications: principles and
  practice}.\hskip 1em plus 0.5em minus 0.4em\relax Prentice Hall PTR New
  Jersey, 1996, vol.~2.

\bibitem{goldsmith2005wireless}
A.~Goldsmith, \emph{Wireless communications}.\hskip 1em plus 0.5em minus
  0.4em\relax Cambridge university press, 2005.

\bibitem{coulson1998statistical}
A.~J. Coulson, A.~G. Williamson, and R.~G. Vaughan, ``A statistical basis for
  lognormal shadowing effects in multipath fading channels,'' \emph{IEEE
  Transactions on Communications}, vol.~46, no.~4, pp. 494--502, 1998.

\bibitem{gudmundson1991correlation}
M.~Gudmundson, ``Correlation model for shadow fading in mobile radio systems,''
  \emph{Electronics letters}, vol.~27, no.~23, pp. 2145--2146, 1991.

\bibitem{shaked1995stochastic}
M.~Shaked, J.~G. Shanthikumar, and Y.~Tong, ``Stochastic orders and their
  applications,'' \emph{SIAM Review}, vol.~37, no.~3, pp. 477--478, 1995.

\bibitem{graziosi2002general}
F.~Graziosi and F.~Santucci, ``A general correlation model for shadow fading in
  mobile radio systems,'' \emph{Communications Letters, IEEE}, vol.~6, no.~3,
  pp. 102--104, 2002.

\bibitem{agrawal2009correlated}
P.~Agrawal and N.~Patwari, ``Correlated link shadow fading in multi-hop
  wireless networks,'' \emph{IEEE Transactions on Wireless Communications},
  vol.~8, no.~8, pp. 4024--4036, 2009.

\bibitem{access2010further}
E.~U. T.~R. Access, ``Further advancements for e-utra physical layer aspects,''
  3GPP TR 36.814, Tech. Rep., 2010.

\bibitem{erceg2004tgn}
V.~Erceg, ``Tgn channel models,'' \emph{IEEE 802.11 document 03/940r4}, 2004.

\bibitem{weber2010overview}
S.~Weber, J.~G. Andrews, and N.~Jindal, ``An overview of the transmission
  capacity of wireless networks,'' \emph{IEEE Transactions on Communications},
  vol.~58, no.~12, pp. 3593--3604, 2010.

\bibitem{baccelli2006aloha}
F.~Baccelli, B.~Blaszczyszyn, and P.~Muhlethaler, ``An aloha protocol for
  multihop mobile wireless networks,'' \emph{IEEE Transactions on Information
  Theory}, vol.~52, no.~2, pp. 421--436, 2006.

\bibitem{zhang2012random}
X.~Zhang and M.~Haenggi, ``Random power control in poisson networks,''
  \emph{Communications, IEEE Transactions on}, vol.~60, no.~9, pp. 2602--2611,
  2012.

\bibitem{zhang2012delay}
------, ``Delay-optimal power control policies,'' \emph{IEEE Transactions on
  Wireless Communications}, vol.~11, no.~10, pp. 3518--3527, 2012.

\bibitem{andrews2011tractable}
J.~G. Andrews, F.~Baccelli, and R.~K. Ganti, ``A tractable approach to coverage
  and rate in cellular networks,'' \emph{IEEE Transactions on Communications},
  vol.~59, no.~11, pp. 3122--3134, 2011.

\bibitem{dhillon2012modeling}
H.~S. Dhillon, R.~K. Ganti, F.~Baccelli, and J.~G. Andrews, ``Modeling and
  analysis of k-tier downlink heterogeneous cellular networks,'' \emph{IEEE
  Journal on Selected Areas in Communications}, vol.~30, no.~3, pp. 550--560,
  2012.

\bibitem{dhillon2012coverage}
H.~S. Dhillon, T.~D. Novlan, and J.~G. Andrews, ``Coverage probability of
  uplink cellular networks,'' in \emph{2012 IEEE Global Communications
  Conference (GLOBECOM)}.\hskip 1em plus 0.5em minus 0.4em\relax IEEE, 2012,
  pp. 2179--2184.

\bibitem{ilow1998analytic}
J.~Ilow and D.~Hatzinakos, ``Analytic alpha-stable noise modeling in a
  {P}oisson field of interferers or scatterers,'' \emph{IEEE Transactions on
  Signal Processing}, vol.~46, no.~6, pp. 1601--1611, 1998.

\bibitem{blaszczyszyn2015wireless}
B.~Blaszczyszyn, M.~K. Karray, and H.~P. Keeler, ``Wireless networks appear
  {P}oissonian due to strong shadowing,'' \emph{IEEE Transactions on Wireless
  Communications}, vol.~14, no.~8, pp. 4379--4390, 2015.

\bibitem{bai2014analysis}
T.~Bai, R.~Vaze, and R.~W. Heath, ``Analysis of blockage effects on urban
  cellular networks,'' \emph{IEEE Transactions on Wireless Communications},
  vol.~13, no.~9, pp. 5070--5083, 2014.

\bibitem{baccelli2015correlated}
F.~Baccelli and X.~Zhang, ``A correlated shadowing model for urban wireless
  networks,'' in \emph{2015 IEEE Conference on Computer Communications
  (INFOCOM)}.\hskip 1em plus 0.5em minus 0.4em\relax IEEE, 2015, pp. 801--809.

\bibitem{zhang2015indoor}
X.~Zhang, F.~Baccelli, and R.~W. Heath, ``An indoor correlated shadowing
  model,'' in \emph{Global Communications Conference (GLOBECOM), 2015
  IEEE}.\hskip 1em plus 0.5em minus 0.4em\relax IEEE, 2015, pp. 1--7.

\bibitem{lee20163}
J.~Lee, X.~Zhang, and F.~Baccelli, ``A 3-d spatial model for in-building
  wireless networks with correlated shadowing,'' \emph{IEEE Transactions on
  Wireless Communications}, vol.~15, no.~11, pp. 7778--7793, 2016.

\bibitem{chiu2013stochastic}
S.~N. Chiu, D.~Stoyan, W.~S. Kendall, and J.~Mecke, \emph{Stochastic geometry
  and its applications}.\hskip 1em plus 0.5em minus 0.4em\relax John Wiley \&
  Sons, 2013.

\bibitem{blaszczyszyn2009directionally}
B.~B{\l}aszczyszyn, D.~Yogeshwaran \emph{et~al.}, ``Directionally convex
  ordering of random measures, shot noise fields, and some applications to
  wireless communications,'' \emph{Advances in Applied Probability}, vol.~41,
  no.~3, pp. 623--646, 2009.

\bibitem{burton1985scaling}
R.~Burton and E.~Waymire, ``Scaling limits for associated random measures,''
  \emph{The Annals of Probability}, pp. 1267--1278, 1985.

\bibitem{pemantle2000towards}
R.~Pemantle, ``Towards a theory of negative dependence,'' \emph{Journal of
  Mathematical Physics}, vol.~41, no.~3, pp. 1371--1390, 2000.

\bibitem{blaszczyszyn2014comparison}
B.~B{\l}aszczyszyn, D.~Yogeshwaran \emph{et~al.}, ``On comparison of clustering
  properties of point processes,'' \emph{Advances in Applied Probability},
  vol.~46, no.~1, pp. 1--20, 2014.

\bibitem{haenggi2012stochastic}
M.~Haenggi, \emph{Stochastic geometry for wireless networks}.\hskip 1em plus
  0.5em minus 0.4em\relax Cambridge University Press, 2012.

\bibitem{alzaid1991laplace}
A.~Alzaid, J.~S. Kim, and F.~Proschan, ``Laplace ordering and its
  applications,'' \emph{Journal of Applied Probability}, vol.~28, no.~1, pp.
  116--130, 1991.

\bibitem{baccelli2009stochasticvol2}
F.~Baccelli and B.~Blaszczyszyn, \emph{Stochastic Geometry and Wireless
  Networks: Volume 2: APPLICATIONS}.\hskip 1em plus 0.5em minus 0.4em\relax Now
  Publishers Inc, 2009, vol.~2.

\bibitem{ganti2009interference}
R.~K. Ganti and M.~Haenggi, ``Interference and outage in clustered wireless ad
  hoc networks,'' \emph{IEEE Transactions on Information Theory}, vol.~55,
  no.~9, pp. 4067--4086, 2009.

\bibitem{baccelli2009stochastic}
F.~Baccelli and B.~Blaszczyszyn, \emph{Stochastic Geometry and Wireless
  Networks: Volume 1: THEORY}.\hskip 1em plus 0.5em minus 0.4em\relax Now
  Publishers Inc, 2009, vol.~1.

\bibitem{daley2007introduction}
D.~J. Daley and D.~Vere-Jones, \emph{An introduction to the theory of point
  processes: volume II: general theory and structure}.\hskip 1em plus 0.5em
  minus 0.4em\relax Springer Science \& Business Media, 2007, vol.~2.

\end{thebibliography}

\end{document}